\documentclass[english,two column, showpacs]{revtex4}
\usepackage[T1]{fontenc}
\usepackage[latin1]{inputenc}
\usepackage{graphicx}
\usepackage{amsmath}
\usepackage{amssymb}
\usepackage{amsfonts}
\begin{document}

\title{Impurity effects on electronic transport in ferropnictide superconductors}
\author{Y.G. Pogorelov,$^1$ M.C. Santos,$^2$ V.M. Loktev$^3$}
\affiliation{$^1$IFIMUP-IN, Departamento de F\'{\i}sica, Universidade do Porto, Porto,
Portugal,\\
$^2$Departamento de F\'{\i}sica, Universidade de Coimbra, R. Larga, Coimbra, 3004-535,
Portugal,\\
$^3$ Bogolyubov Institute for Theoretical Physics, NAN of Ukraine, 14b Metrologichna str.,
03143 Kiev, Ukraine}

\begin{abstract}
Effects of impurities and disorder on transport properties by electronic quasiparticles in
superconducting iron pnictides are theoretically considered. The most prominent new features
compared to the case of pure material should appear at high enough impurity concentration
when a specific narrow band of conducting quasiparticle states can develop within the
superconducting gap, around the position of localized impurity level by a single impurity
center. The predicted specific threshold effects in the frequency dependent optical conductivity
and temperature dependent thermal conductivity and also in Seebeck and Peltier coefficients 
can have interesting potentialities for practical applications.
\end{abstract}

\maketitle\

\section{Introduction}

A considerable interest in actual research of superconductivity (SC) with high critical temperature 
is focused on the family of doped ferropnictide compounds \cite{kamihara1, kamihara2} and one 
of their notable distinctions from "old" BCS superconductors and more recent doped perovskite 
systems consists in possibility for a peculiar, so-called extended s-wave symmetry of superconducting
order parameter which changes its sign between electron and hole segments of the Fermi surface
\cite{mazin}. This additional property permits to avoid the fundamental limitation by the Anderson's 
theorem \cite{and} for non-magnetic impurities to produce localized impurity levels within the 
superconducting band gap \cite{dzhang, zhang}. At finite, but low enough, impurity concentration, 
such levels are expected to give rise to some resonance effects like those well studied in 
semiconductors at low doping concentrations \cite{shklo}. Analogous effects in superconductors 
were theoretically predicted and experimentally discovered for magnetic impurities, either in BCS 
systems \cite{shiba,rus,maki} and in the two-band MgB$_2$ system \cite{con, moca}. In all those 
cases, the breakdown of the Anderson's theorem is only due to the breakdown of the spin-singlet 
symmetry of an $s$-wave Cooper pair by a spin-polarized impurity, and the main physical interest 
of the considered case of SC iron pnictides from the point of view of disorder in general is the 
possibility for pair-breaking even on non-magnetic impurity scatterers. The latter theoretical prediction 
was confirmed by the observations of various effects from localized impurity states, for instance, in the 
superfluid density (observed through the London penetration length) \cite{gordon, kitagawa}, transition 
critical temperature \cite{guo, li} and electronic specific heat \cite{hardy}, all mainly due to an emerging 
spike of electronic density of states against its zero value in the initial band gap.

But it is also known that indirect interactions between random impurity centers of certain type (the 
so-called deep levels at high enough concentrations) in doped semiconductors can lead to formation 
of collective band-like states \cite{iv, yama}. This corresponds to the Anderson transition in a general 
disordered system \cite{and1}, and the emerging new band of quasiparticles in the spectrum can 
essentially change thermodynamics and transport in the doped material \cite{mott}. An intriguing 
possibility for similar banding of impurity levels within the SC gap \cite{bal, lokt} was recently discussed 
for the doped ferropnictides \cite{pog}. The present work is aimed on a more detailed analysis of the 
band-like impurity states, focused on their observable effects that cannot be produced by localized 
impurity states. We use the specific form of Green functions for superconducting quasiparticles 
derived in the previous work \cite{pog} in the general Kubo-Greenwood formalism \cite{kubo} to obtain 
the temperature and frequency dependences of optical and thermal conductivity and also of thermoelectric 
coefficients. These results are compared with the available experimental data and some suggestions are 
done on possible practical applications of such impurity effects.

\section{Green functions for disordered SC ferropnictide}
We begin from a brief summary of the Green function description of electronic spectrum in
LaOFeAs with impurities (not necessarily dopants) using the minimal coupling model 
\cite{dag,tsai} for the non-perturbed Hamiltonian. It considers only 2 types of local Fe 
orbitals, $d_{xz}$ (or $x$) and $d_{yz}$ (or $y$), on sites of square lattice with lattice 
parameter $a$ and 4 hopping parameters between nearest neighbors (NNs) and next nearest 
neighbors (NNNs): i) $t_1$ for $xx$ or $yy$ NNs along their orientations, and $t_2$ across 
them, and ii) $t_3$ for $xx$ or $yy$ NNNs, and $t_4$ for $xy$ NNNs. The resulting band 
Hamiltonian is diagonal in quasimomentum ${\bf k}$ and spin $\sigma$, but non-diagonal with respect 
to the orbital indices of the 2-spinors $\psi^\dagger({\bf k},\sigma) = (x^\dagger_{{\bf k},\sigma},
y^\dagger_{{\bf k},\sigma})$:
\begin{equation}
 H_t = \sum_{{\bf k},\sigma} \psi^\dagger({\bf k},\sigma)\hat h({\bf k})\psi({\bf k},\sigma).
 \label{eq1}
  \end{equation}
Here the energy matrix in orbital basis is expanded in Pauli matrices $\hat\sigma_i$: 
$\hat h({\bf k}) = \varepsilon_{+,{\bf k}}\hat\sigma_0 + \varepsilon_{-,{\bf k}}\hat\sigma_3 + \varepsilon_{xy,{\bf k}}
\hat\sigma_1$ with the energy factors $\varepsilon_{\pm,{\bf k}} = (\varepsilon_{x,{\bf k}} \pm \varepsilon_{x,{\bf k}})/2$, 
and
\begin{eqnarray}
 \varepsilon_{x,{\bf k}} & = & - 2t_1 \cos ak_x - 2t_2 \cos ak_y - 4t_3 \cos ak_x \cos ak_y,\nonumber\\
  \varepsilon_{y,{\bf k}} & = & - 2t_1 \cos ak_y - 2t_2 \cos ak_x - 4t_3 \cos ak_x \cos ak_y,\nonumber\\
   \varepsilon_{xy,{\bf k}} & = & - 4t_4 \sin ak_x \sin ak_y.\nonumber
    \end{eqnarray}
It is readily diagonalized at passing from the orbital to subband basis: $ \hat h_b({\bf k})
= \hat U({\bf k})\hat h({\bf k})\hat U({\bf k})^\dagger$, with the unitary matrix $\hat U({\bf k}) =
\exp(-i\hat\sigma_2\theta_{\bf k}/2)$ and $\theta_{\bf k} = \arctan \left(\varepsilon_{xy,{\bf k}}/\varepsilon_{-,{\bf k}}\right)$.
The resulting eigen-energies for electron and hole subbands are:
\begin{equation}
 \varepsilon_{h,e}({\bf k}) = \varepsilon_{+,{\bf k}} \pm \sqrt{\varepsilon_{xy,{\bf k}}^2 + \varepsilon_{-,{\bf k}}^2},
 \label{eq2}
  \end{equation}
and respective electron and hole segments of the Fermi surface are defined by the equations
$\varepsilon_{e,h}({\bf k}) = \varepsilon_{\rm F}$. A reasonable fit to the  LaOFeAs band structure by the more
detailed LDA calculations \cite{xu} is attained with the parameter choice (in $|t_1|$ units)
of $t_1 = -1$, $t_2 = 1.3$, $t_3 = t_4 = -0.85$ \cite{raghu}.

The SC state of such multiband electronic system is suitably described in terms of "multiband
-Nambu" 4-spinors $\Psi_{\bf k}^\dagger = \left(\alpha_{{\bf k},\uparrow}^\dagger,\alpha_{-{\bf k},\downarrow}, \beta_{{\bf k},\uparrow}
^\dagger,\beta_{-{\bf k},\downarrow}\right)$ with the multiband spinor $\left(\alpha_{{\bf k},\sigma}^\dagger,\beta_{{\bf k},\sigma}
^\dagger\right) = \psi^\dagger({\bf k},\sigma)\hat U^\dagger({\bf k})$, by a 4$\times$4 extension of the
Hamiltonian Eq. \ref{eq1} in the form:
\begin{equation}
 H_s = \sum_{{\bf k},\sigma} \Psi_{\bf k}^\dagger\hat h_s({\bf k})\Psi_{\bf k},
 \label{eq3}
  \end{equation}
where the 4$\times$4 matrix $\hat h_s({\bf k}) = \hat h_b({\bf k})\otimes\hat\tau_3 + \Delta_{{\bf k}}\hat\sigma_0
\otimes\hat\tau_1$, includes the Pauli matrices $\hat\tau_i$ acting on the Nambu (particle-antiparticle)
indices in $\Psi$-spinors. The simplified form for the extended \emph{s}-wave gap function takes
constant values, $\Delta_{\bf k} = \Delta$ on the electron segments and $\Delta_{\bf k} = -\Delta$ on the hole segments.

The observable values result from the (Fourier transformed) GF 4$\times$4 matrices $\hat
G_{{\bf k},{\bf k}'} = \langle\langle\Psi_{{\bf k}}|\Psi_{{\bf k}'}^\dagger\rangle\rangle$, and for the
non-perturbed system, Eq. \ref{eq1}, they are diagonal in quasimomentum: $\hat G_{{\bf k},{\bf k}'}
= \delta_{{\bf k},{\bf k}'} \hat g_{\bf k}$ with
\begin{eqnarray}
 \hat g_{\bf k} & = & \frac{\varepsilon\hat\tau_0 + \varepsilon_e({\bf k})
 \hat\tau_3 + \Delta\hat\tau_1}{2d_{e,{\bf k}}}\otimes\hat\sigma_e\nonumber\\
 & + & \frac{\varepsilon\hat\tau_0 + \varepsilon_h({\bf k})\hat\tau_3 - \Delta\hat\tau_1}{2d_{h,{\bf k}}}\otimes\hat\sigma_h,
  \label{eq4}
   \end{eqnarray}
$\hat\sigma_{e,h} = \left(\hat\sigma_0 \pm \hat\sigma_3\right)/2$, $d_{i,{\bf k}} = \varepsilon^2 - \varepsilon_i^2({\bf k}) - \Delta^2$.

To simplify the treatment of impurity perturbations, the band structure is approximated to 
identical circular electron and hole Fermi segments of radius $k_{\rm F}$ around respective 
points ${\bf K}_{e,h}$ in the Brillouin zone and to similar linear dispersion of normal state 
quasiparticles near the Fermi level $\varepsilon_{\rm F}$: $\varepsilon_e({\bf k}) -  \varepsilon_{\rm F} = \hbar v_{\rm F}
\left(|{\bf k} - {\bf K}_e| - k_{\rm F}\right)$ and $\varepsilon_h({\bf k}) -  \varepsilon_{\rm F} = -\hbar v_{\rm F} 
\left(|{\bf k} - {\bf K}_h| - k_{\rm F}\right)$. Moreover, we shall describe the contributions of 
both segments to overall electronic properties by a single quasimomentum variable $\xi$ that 
identifies electron $\xi_e = \varepsilon_e({\bf k}) -  \varepsilon_{\rm F}$ and hole $\xi_h = \varepsilon_h({\bf k}) -  \varepsilon_{\rm F}$ 
ones.

Next, the Hamiltonian of the disordered SC system is chosen as $H = H_s + H_{imp}$ including 
besides $H_s $, Eq. \ref{eq3}, the term  due to non-magnetic impurities \cite{dzhang}  on random 
sites ${\bf p}$ in Fe square lattice with an on-site energy shift $V$ (supposed positive without loss 
of generality). It is written in the multiband-Nambu spinor form as:
\begin{equation}
H_{imp} = {1 \over N} \sum_{{\bf p},{\bf k},{\bf k}'} {\rm e}^{i({\bf k}' - {\bf k})\cdot{\bf p}}\Psi_{\bf k}^\dagger 
\hat V_{{\bf k},{\bf k}'}\Psi_{{\bf k}'},
 \label{eq5}
   \end{equation}
with the number $N$ of unit cells in the crystal and the 4$\times$4 scattering matrix 
$\hat V_{{\bf k},{\bf k}'} = V\hat U^\dagger({\bf k})\hat U({\bf k}')\otimes\hat\tau_3$. In presence of this 
perturbation, the GFs can be expressed in specific forms depending on whether the considered 
quasiparticle energy falls into the range of band-like or localized states. Namely, for band-like 
states, the momentum diagonal GF:
\begin{equation}
 \hat G_{\bf k} = \hat G_{{\bf k},{\bf k}} = (\hat g_{\bf k}^{-1} - \hat \Sigma_{\bf k})^{-1},
  \label{eq6}
   \end{equation}
involves the self-energy matrix $\hat \Sigma_{\bf k}$  in the form of the so-called renormalized 
group expansion \cite{ilp}:
\begin{equation}
 \hat \Sigma_{\bf k} = c\hat T \left(1 + c \hat B_{\bf k} + \dots\right).
  \label{eq7}
    \end{equation} 
This series in powers of impurity concentration $c$ begins from the (k-independent) T-matrix, 
$\hat T = \hat V \left(1 - \hat G \hat V\right)^{-1}$. From the matrices $\hat V = 
\hat V_{{\bf k},{\bf k}} = V\hat\tau_3$ and $\hat G = N^{-1} \sum_{\bf k} \hat g_{\bf k} = \pi \varepsilon \rho_{\rm F}
\hat\tau_0/\sqrt{\Delta^2 - \varepsilon^2}$ (with the Fermi density of states $\rho_{\rm F}$ and the henceforth 
omitted trivial factor $\hat\sigma_0$), the T-matrix explicit form is:
\begin{equation}
 \hat T = \frac{V}{1 + v^2}\frac{v\varepsilon\sqrt{\Delta^2 - \varepsilon^2}\hat\tau_0 - \left(\Delta^2 - \varepsilon^2\right)
  \hat\tau_3}{\varepsilon^2 - \varepsilon_0^2}.
   \label{eq8}
     \end{equation} 
where $\varepsilon_0 = \Delta/\sqrt{1 + v^2}$ defines the in-gap impurity levels \cite{tsai} through the 
dimensionless impurity perturbation parameter $v = \pi\rho_{\rm F}V$. Inside the gap, the 
T-matrix, Eq. \ref{eq8},  is a real function which can be approximated near the impurity 
levels $\pm\varepsilon_0$ as: $\hat T \approx \gamma^2 \left(\varepsilon  \hat\tau_0 - \varepsilon_0 \hat\tau_3\right)/\left(\varepsilon^2 
- \varepsilon_0^2\right)$, with the effective coupling constant  $\gamma^2 = V\varepsilon_0(v\varepsilon_0/\Delta)^2$. In contrary, 
outside the gap it is dominated by its imaginary part: Im$\hat T = (\gamma^2/v\varepsilon_0) \varepsilon \sqrt{\varepsilon^2 - 
\Delta^2}/\left(\varepsilon^2 - \varepsilon_0^2\right)$.

The next terms besides unity in the brackets of Eq. \ref{eq7} describe the effects of indirect 
interactions between impurities, with $\hat B_{\bf k} $ related to pairs and the omitted terms to 
groups of three and more impurities. The series convergence defines the energy ranges of 
band-like states, delimited by the Mott mobility edges $\varepsilon_c$ \cite{mott}. Within the band-like 
energy ranges, the self-energy matrix can be safely approximated by the T-matrix, $\hat \Sigma_{\bf k} 
\approx c \hat T$, and the dispersion laws for corresponding bands at given quasimomentum 
${\bf k}$ are defined from the $\hat G_{\bf k}$ denominator:
\begin{eqnarray}
 D_{\bf k}(\varepsilon) & = & \det \hat G_{\bf k}^{-1}(\varepsilon) = \tilde d_{e,{\bf k}}(\varepsilon) \tilde d_{h,{\bf k}}(\varepsilon) \nonumber\\
  & = & \left(\tilde\varepsilon^2 - \tilde\xi_e^2 - \Delta^2\right)\left(\tilde\varepsilon^2 - \tilde\xi_h^2 - 
   \Delta^2\right),
    \label{eq9}
     \end{eqnarray}
with the renormalized energy and momenta forms:
\begin{eqnarray}
  \tilde\varepsilon & = & \varepsilon\left(1 - \frac{c V v}{1 + v^2}\frac{\sqrt{\Delta^2 - \varepsilon^2}}{\varepsilon^2 -  \varepsilon_0^2}\right),\nonumber\\
&& \tilde\xi_j = \xi_j -  \frac{c V}{1 + v^2} \frac{\Delta^2 - \varepsilon^2}{\varepsilon^2 - \varepsilon_0^2}.\nonumber
    \end{eqnarray}
The roots of the dispersion equation Re $D_{\bf k}(\varepsilon) = 0$ define up to 8 subbands:  4 of them 
with energies near the roots of the non-perturbed denominators $d_{j,{\bf k}}$ in the $e$- and 
$h$-segments can be called "principal" or $pr$-bands, they are similar to quasiparticles in 
the pure crystal; and other 4, "impurity" or $imp$-bands, with energies near $\pm \varepsilon_0$ in the 
same segments are only specific for disordered systems. The dispersion law for $p$-bands
is presented in the $\xi$-scale as:
\begin{equation}
  \varepsilon_{pr}(\xi) \approx \sqrt{\xi^2  + \Delta^2},
    \label{eq10}
      \end{equation} 
and it only differs from the non-perturbed one by the finite linewidth $\Gamma(\varepsilon) \approx c{\rm Im}\hat T$, 
so that the validity range of Eq. \ref{eq10} defined from the known Ioffe-Regel-Mott criterion, 
$\xi d\varepsilon_b/d\xi \gtrsim \Gamma(\varepsilon_b(\xi))$ \cite{IR}, \cite{mott} as $\xi \gtrsim c/(\pi\rho_{\rm F})$. This defines 
the mobility edge in closeness to the gap edge,
\begin{equation}
  \varepsilon_c - \Delta \sim c^2/c_0^{4/3}\Delta.
    \label{eq11}
      \end{equation} 
Here $c_0 = (\pi\rho_{\rm F}\varepsilon_0)^{3/2}/\left(a k_{\rm F}\right)\sqrt{2v/(1 + v^2)}$  is the characteristic 
impurity concentration such that the impurity bands emerge just at $c > c_0$ \cite{pog}. Their dispersion 
(in $\xi$) for the exemplar case of positive energies and $e$-segment is approximated as:
\begin{equation}
 \varepsilon_{imp}(\xi) \approx \varepsilon_0 + c\gamma^2\frac{\xi  - \varepsilon_0}{\xi^2 + \xi_0^2}.
   \label{eq12}
     \end{equation} 
   
\begin{figure}[htbp]
\begin{center}
\includegraphics[width=8cm]{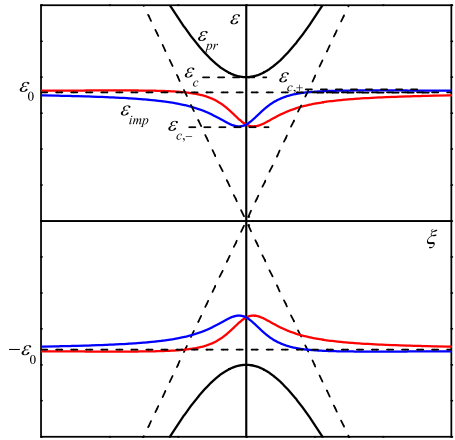}
\caption{Dispersion laws in the modified quasiparticle spectrum of a SC ferropnictide with 
impurities. The impurity perturbation parameters were chosen as: $v = 0.5$, $c_0 = 1.3\cdot 
10^{-3}$, $c_1 =  1.7\cdot 10^{-2}$, $c = 4\cdot 10^{-3}$. For compactness, the plot superimposes 
the blue lines for the in-gap impurity subbands near the electron-like pockets of the Fermi surface 
and red lines for those near the hole-like pockets.}
\label{fig1}
\end{center}
\end{figure}

The formal upper limit energy by Eq. \ref{eq12}, $\varepsilon_+ = \varepsilon_0 + c\gamma^2/[2(\Delta 
+ \varepsilon_0)]$, is attained at $\xi = \xi_+ = \varepsilon_0 +  \Delta$ and the lower limit $\varepsilon_- 
= \varepsilon_0 - c\gamma^2/[2(\Delta - \varepsilon_0)]$ at $\xi_- = \varepsilon_0 - \Delta$. But in fact, 
this dispersion law is only valid until the related mobility edges $\varepsilon_{c,\pm}$ whose onset 
near the $i$-band edges is due to the higher terms in the group expansion, Eq. \ref{eq7}, and amounts 
to:
\begin{eqnarray}
\varepsilon_+ -  \varepsilon_{c,+} & \sim & \left(\varepsilon_{max} -  \varepsilon_0\right)\left(\frac{c_0}c\right)^4,\nonumber\\
  \varepsilon_{c,-} - \varepsilon_-  & \sim & \left(\varepsilon_0 -  \varepsilon_{min}\right)\left(\frac{c_0}c\right)^4.
 \label{eq13}
  \end{eqnarray}
These limitations restrict $\xi$ to beyond some vicinities of the extremal points: $|\xi - \xi_\pm| \gtrsim 
\xi_\pm\left(c_0/c\right)^2$ (narrow enough at $c \gg c_0$). Another limitation is that $\xi$ not be too 
far from these points: $|\xi - \xi_\pm| \lesssim \xi_\pm(c/c_0)^4$. A symmetric replica of Eq. \ref{eq12} 
near $-\varepsilon_0$ at the $e$-segment is the impurity subband with the dispersion law $-\varepsilon_i(\xi)$. Yet two 
more impurity subbands near the $h$-segment are described in the unified $\xi$ frame by the inverted 
dispersion laws $\pm\varepsilon_{imp}(-\xi)$. The overall composition of band-like states in this frame is shown in Fig. 
\ref{fig1}. It is also important to notice that the above described in-gap impurity band structure is only 
justified until it is narrow enough compared to the SC gap $\Delta$ itself. From Eq. \ref{eq12}, this requires 
that the impurity concentration stays well below the upper critical value
\[c_1 = \pi\rho_{\rm F}\Delta\sqrt{1 + v^2}.\]
that can amount about few percents. In what follows, the condition $c \ll c_1$ is presumed.

At least, for $c < c_0$, all the in-gap states are localized and more adequately described by 
an alternative, the so called non-renormalized group expansion of $\hat G_{\bf k}$ (though this 
case is beyond the scope of the present study) while the principal bands are still defined by 
Eqs. \ref{eq10},\ref{eq11}.

In-gap impurity states, either localized and band-like,  can produce notable resonance effects 
on various thermodynamical properties of disordered superconductors, as transition critical 
temperature, London penetration length, electronic specific heat, etc. \cite{pog}. But besides that, 
other effects, only specific for new quasiparticle bands, can be expected on kinetic properties of 
the disordered material, while the localized impurity states should have practically no effect on 
them. Such phenomena can be naturally described in terms of the above indicated GF matrices as 
seen in what follows.

\section{Kubo-Greenwood formalism for multiband superconductor}

The relevant kinetic coefficients for electronic processes in the considered disordered superconductor 
follow from the general Kubo-Greenwood formulation \cite{kubo}, adapted here to the specific multiband 
structure of Green function matrices. Thus, one of the basic transport characteristics, the (frequency and 
temperature dependent) electrical conductivity is expressed in this approach as:
\begin{eqnarray}
\sigma(\omega,T)  & = & \frac{e^2 }{\pi} \int d\varepsilon\frac{f(\varepsilon) - f(\varepsilon')}\omega \int d{\bf k} v_x({\bf k},\varepsilon) 
 v_x({\bf k},\varepsilon') \nonumber\\
  & \times & {\rm Tr}\left[{\rm Im}\hat G_{{\bf k}}(\varepsilon){\rm Im}\hat G_{{\bf k}}(\varepsilon')\right],
   \label{eq14}
    \end{eqnarray}
for $\varepsilon' = \varepsilon - \hbar\omega$ and the electric field applied along the $x$-axis. Besides the common Fermi 
occupation function $f(\varepsilon) = ({\rm e}^{\beta\varepsilon} + 1)^{-1}$ with the inverse temperature $\beta = 1/k_{\rm B}T$, the 
above formula involves the generalized velocity function:
\begin{equation}
 {\bf v}({\bf k},\varepsilon) = \left(\hbar\frac{\partial {\rm Re}D_{\bf k}(\varepsilon)}{\partial \varepsilon}\right)^{-1}{\bf \nabla}_{\bf k} 
  {\rm Re}D_{\bf k}(\varepsilon).
   \label{eq15}
    \end{equation} 
This function is defined in the whole $\xi,\varepsilon$ plane in a way to coincide with the physical quasiparticle 
velocities for each particular band, Eqs. \ref{eq9}, \ref{eq12}, along the corresponding dispersion laws: 
${\bf v}({\bf k},\varepsilon_j({\bf k})) = \hbar^{-1}{\bf \nabla}_{\bf k} \varepsilon_j({\bf k}) = v_{j,{\bf k}}$, $j = pr,imp$. The conductivity resulting 
from Eq. \ref{eq13} can be then used for calculation of optical reflectivity.

Other relevant quantities are the static (but temperature dependent) transport coefficients, as the heat 
conductivity:
\begin{eqnarray}
\kappa(T) & = & \frac{\hbar}{\pi } \int d\varepsilon \frac{\partial f(\varepsilon)}{\partial \varepsilon} \varepsilon^2 \int d{\bf k} 
 \left[v_x({\bf k},\varepsilon)\right]^2 \nonumber\\
& \times & {\rm Tr}\left[{\rm Im}\hat G_{{\bf k}}(\varepsilon)\right]^2,
 \label{eq16}
  \end{eqnarray}
and the thermoelectric coefficients associated with the static electrical conductivity $\sigma(T) \equiv \sigma(0,T)$ 
\cite{note}, the Peltier coefficient:
\begin{eqnarray}
 \Pi(T) & = & \frac{\hbar e}{\pi \sigma(0,T)} \int d\varepsilon \frac{\partial f(\varepsilon)}{\partial \varepsilon} \varepsilon \int 
  d{\bf k} \left[v_x({\bf k},\varepsilon)\right]^2 \nonumber\\
& \times & {\rm Tr}\left[{\rm Im}\hat G_{{\bf k}}(\varepsilon)\right]^2,
 \label{eq17}
  \end{eqnarray}
and the Seebeck coefficient $S(T) = \Pi(T)/T$.  All these transport characteristics, though being relatively 
more complicated from the theoretical   point of view than the purely thermodynamical quantities as, e.g., 
specific heat or London penetration length \cite{pog}, permit an easier and more reliable experimental 
verification and so could be of higher interest for practical applications of the considered impurity effects 
in the multiband superconductors.

Next, we consider the particular calculation algorithms for the expressions, Eqs.  \ref{eq14}, 
\ref{eq16},  \ref{eq17}, beginning from the more involved case of dynamical conductivity, Eq.  
\ref{eq14}, and then reducing it to simpler static quantities, Eqs.  \ref{eq16},  \ref{eq17}.

\section{Optical conductivity}

\begin{figure}[htbp]
\begin{center}
\includegraphics[width=8cm]{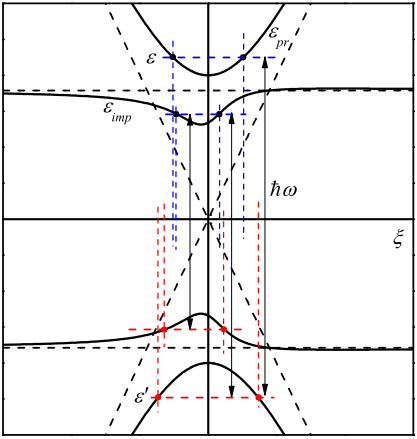}
\caption{Configuration of the poles $\xi_j$ of GF's contributing to different types of optical conductivity 
processes over one part (electronic pocket) of the quasiparticles spectrum by Fig. \ref{fig1}.}
\label{fig2}
\end{center}
\end{figure}

The integral in Eq. \ref{eq14} is dominated by the contributions from $\delta$-like peaks of the 
${\rm Im}\hat G_{{\bf k}}(\varepsilon)$ and ${\rm Im}\hat G_{{\bf k}}(\varepsilon ')$ matrix elements. 
These peaks arise from the above dispersion laws, Eqs. \ref{eq9}, \ref{eq11}, thus restricting the 
energy integration to the band-like ranges: $|\varepsilon| > \varepsilon_c$ for the $pr$-bands and 
$\varepsilon_{c,-} < |\varepsilon| < \varepsilon_{c,+}$ for the $imp$-bands. Regarding the occupation 
numbers $f(\varepsilon)$ and $ f(\varepsilon')$ at reasonably low temperatures $k_{\rm B}T \ll 
\Delta,\varepsilon_0$, the most effective contributions correspond to positive $\varepsilon$ values, 
either from $pr$- or $imp$-bands, and to negative $\varepsilon'$ values from their negative counterparts, 
$pr'$ or $imp'$. There are three general kinds of such contributions: i) $pr-pr'$, due to transitions between 
the principal bands, similar to those in optical conductivity by the pure crystal (but with a slightly shifted 
frequency threshold: $\hbar\omega \geq 2\varepsilon_c$), ii) $pr-imp'$ (or $imp-pr'$), due to combined 
transitions between the principal and impurity bands within the frequency range $\hbar\omega \geq 
\varepsilon_c + \varepsilon_{c,-}$, and iii) $imp-imp'$, due to transitions between the impurity bands within 
a narrow frequency range of $2\varepsilon_{c,-} < \hbar\omega < 2\varepsilon_{c,+}$. The frequency-momentum 
relations for these processes and corresponding peaks are displayed in Fig. \ref{fig2}. The resulting optical 
conductivity reads $\sigma(\omega,T) = \sum_\nu \sigma_\nu(\omega,T)$ with $\nu = pr-pr',\,imp-imp'$, and 
$imp-pr'$.

For practical calculation of each contribution, the relevant matrix Im$\hat G_{\bf k}(\varepsilon)$ (within 
the band-like energy ranges) can be presented as Im$\hat G_{\bf k}(\varepsilon) = \hat N(\varepsilon,\xi) {\rm Im}
\left[D_{\bf k}(\varepsilon)^{-1}\right]$ where the numerator matrix:
\begin{equation}
 \hat N(\varepsilon,\xi) = {\rm Re}\left(\tilde\varepsilon + \tilde\xi \hat \tau_3 +  \Delta \hat \tau_1\right),
   \label{eq18}
     \end{equation} 
is a smooth enough function while the above referred peaks result from zeros of Re$D_{\bf k}(\varepsilon)$. 
Now, the quasimomentum integration in Eq. \ref{eq14} under the above chosen symmetry of 
Fermi segments spells as $\int d{\bf k} = 2(h v_{\rm F})^{-1}\int d\varphi \int d\xi$ where the factor 2 
accounts for identical contributions from $e$- and $h$-segments. The azimuthal integration 
contributes by the factor of $\pi$ (from $x$-projections of velocities) and the most important radial 
integration is readily done after expanding its integrand in particular pole terms:
\begin{eqnarray}
  v(\xi,\varepsilon)v(\xi,\varepsilon') & {\rm Tr} & \left[{\rm Im}\hat G(\xi,\varepsilon){\rm Im}\hat G(\xi,\varepsilon')\right] 
   \nonumber\\
& = & \sum_\alpha A_\alpha(\varepsilon,\varepsilon') \delta\left(\xi - \xi_\alpha\right),
 \label{eq19}
  \end{eqnarray}
 where $v(\xi,\varepsilon) = |{\bf v}({\bf k},\varepsilon)|$ and $\hat G(\xi,\varepsilon') \equiv \hat G_{\bf k}(\varepsilon')$ 
 define the  respective  residues:
 \begin{equation}
   A_\alpha(\varepsilon,\varepsilon') = \pi v_\alpha v'_\alpha \frac{\tilde\varepsilon\tilde\varepsilon' + \tilde\xi\tilde\xi' + 
     \Delta^2}{\prod_{\beta \neq \alpha}\left(\xi_\alpha - \xi_\beta\right)}.
       \label{eq20}
         \end{equation} 

Here $v_\alpha \equiv v\left(\varepsilon,\xi_\alpha\right)$, $v'_\alpha \equiv v\left(\varepsilon',\xi_\alpha\right)$, and 
the indices $\alpha,\beta$ run over all the poles  of the two Green  functions.  As follows from Eqs.  \ref{eq10}, \ref{eq12} 
and seen in Fig. \ref{eq2}, there can be two such poles of $\hat G(\xi,\varepsilon)$ related to band-like states with 
positive $\varepsilon$ and respective quasi-momentum values denoted as $\xi_{1,2}(\varepsilon)$. For energies 
within the $pr$-band, $\varepsilon > \varepsilon_c$, they are symmetrical: 
\begin{equation}
  \xi_{1,2}(\varepsilon) \approx \pm  \sqrt{\varepsilon^2 - \Delta^2},
    \label{eq21}
      \end{equation} 
while within the $imp$-band at $\varepsilon_{c,-} < \varepsilon < \varepsilon_{c,+}$, their positions are asymmetrical:
\begin{equation}
  \xi_{1,2}(\varepsilon) \approx \frac{c\gamma^2 \mp 2\varepsilon_0\sqrt{\left(\varepsilon_+ - \varepsilon\right)
    \left(\varepsilon - \varepsilon_-\right)}}{2\left(\varepsilon - \varepsilon_0\right)}.
       \label{eq22}
         \end{equation} 
Notice also that, within the $imp$-band, there is a narrow vicinity of $\varepsilon_0$ of $\sim c_0^{1/3} (c_0/c)^3 \varepsilon_0$ 
width where only the $\xi_1$ pole by Eq. \ref{eq22} is meaningful and the other contradicts to the IRM criterion (so that there is 
no band-like states with that formal $\xi_2$ values in this energy range).  Analogous poles of $\hat G(\xi,\varepsilon')$ at negative 
$\varepsilon'$ are referred to as $\xi_{3,4}(\varepsilon')$ in what follows. Taking into account a non-zero Im$D_{\bf k}(\varepsilon)$ 
(for the $imp$-band, it is due to the non-trivial terms in the group expansion, Eq. \ref{eq7}), each $\alpha$th pole becomes a 
$\delta$-like peak with an effective linewidth $\Gamma_\alpha$ but this value turns to be essential (and will be specified) only at 
calculation of static coefficients like Eqs. \ref{eq16}, \ref{eq17}.

Since four peaks in Eq. \ref{eq19} for optical conductivity are typically well separated, the $\xi$-integration 
is trivially done considering them true $\delta$-functions, then the particular terms in $\sigma(\omega,T)$ follow as the 
energy integrals:
\begin{equation}
  \sigma_\nu(\omega,T) = 2e^2 \int_{\varepsilon_{\nu,-}}^{\varepsilon_{\nu,+}} d\varepsilon\frac{f(\varepsilon) - f(\varepsilon')}
    \omega \sum_{\alpha=1}^4 A_\alpha(\varepsilon,\varepsilon'),
      \label{eq23}
        \end{equation} 
where $\nu$ takes the values $pr-pr'$, $imp-pr'$, or $imp-imp'$ and the limits $\varepsilon_{\nu,\pm}$ should assure that both 
$\varepsilon$ and $\varepsilon'$ are kept within the respective band-like energy ranges. 

Thus, in the $pr-pr'$ term, the symmetry of the poles $\xi_{1,2}(\varepsilon)$ and $\xi_{3,4}(\varepsilon')$ by Eq. \ref{eq21} and the
symmetry of $pr$- and $pr'$-bands themselves defines their equal contributions, then using simplicity of the generalized velocity 
function $v(\xi,\varepsilon) = \xi/\varepsilon$ and  the non-renormalized energy and momentum variables, $\tilde\varepsilon \to 
\varepsilon$, $\tilde\xi \to \xi$, the energy integration between the limits $\varepsilon_{pr-pr',-} = \varepsilon_c$ and 
$\varepsilon_{pr-pr',+} =  \hbar\omega - \varepsilon_c$ provides its explicit analytic form as $\sigma_{pr-pr'}(\omega,T) = 
\sigma_{pr-pr'}(\omega,0) - \sigma_{pr-pr',T}(\omega)$. Here the zero-temperature limit value is:
\begin{widetext}
\begin{eqnarray}
  \sigma_{pr-pr'}(\omega,0) & \approx & \sigma_0\frac{2\omega_c}{\omega^2} \left\{\sqrt{4\omega^2 - \omega_c^2} \ln\left[ 2 
    \frac{\omega(2\omega - \omega_c) +  \sqrt{\omega(\omega - \omega_c)    (4\omega^2 - \omega_c^2)}}{\omega_c^2} - 1\right] 
      \right.\nonumber\\
& + &  \left. 2\omega \ln\left[2\frac{\omega - \sqrt{\omega(\omega - \omega_c)}}{\omega_c} - 1\right] - 2\sqrt{\omega(\omega - 
    \omega_c)} \right\},
        \label{eq24}
          \end{eqnarray}
\end{widetext}
with the characteristic scale $\sigma_0 = e^2/\Delta^2$ and simple asymptotics: 
\begin{eqnarray}
  \sigma_{pr-pr'}(\omega,0) & \approx & (2/3)\sigma_0(\omega /\omega_c -1 )^{3/2}, \quad \omega - \omega_c \ll \omega_c,\nonumber\\
 \sigma_{pr-pr'}(\omega,0) & \approx & \sigma_0(32\omega_c/\omega)\ln(2\omega/\omega_c), \quad\quad\quad \omega \gg 
   \omega_c,\nonumber
     \end{eqnarray} 
with respect to the threshold frequency $\omega_c = 2\varepsilon_c/\hbar$, reaching the maximum value $\approx 1.19\sigma_0$ 
at $\omega \approx 2.12\omega_c$ as seen in Fig. 3. The (small) finite-temperature correction to the above value:
\begin{widetext}
\begin{eqnarray}
 \sigma_{pr-pr',T}(\omega) \approx  \sigma_0\frac{2\omega_c^2{\rm e}^{-\beta\Delta}} {\beta\hbar(\omega - \omega_c)\omega\sqrt\Delta}
   &&\left[\frac{\sqrt{\hbar\omega}} \Delta \left(1 - \frac{F(\sqrt{\beta\hbar(\omega - \omega_c)})} {\sqrt{\beta\hbar(\omega - \omega_c)}}\right) 
     \right.\nonumber\\
 && \qquad\qquad +  \left.\frac{\sqrt{2\Delta}}{\hbar\omega - \Delta}\left(\frac{\sqrt \pi}2 \frac{{\rm erf} (\sqrt{\beta\hbar(\omega - \omega_c)})}
    {\sqrt{\beta\hbar(\omega - \omega_c)}} - {\rm e}^{-\beta\hbar(\omega - \omega_c)}\right)\right],
       \label{eq25}
         \end{eqnarray}
\end{widetext}
involves the Dawson function $F(z) = \sqrt\pi {\rm e}^{-z^2} {\rm erf}(iz)/(2i)$ and the error function ${\rm erf}(z)$ \cite{abst}. 

Calculation of the $imp-pr'$-term is more complicated since asymmetry of the $imp$-band poles $\xi_{1,2}(\varepsilon)$ by Eq. 
\ref{eq22} and their non-equivalence to the symmetric poles $\xi_{3,4}(\varepsilon')$ of the $pr'$-band analogous to Eq. 
\ref{eq21}. More complicated expressions also define the generalized velocity function within the $imp$-band range: 
\begin{equation}
  \hbar v(\xi,\varepsilon) = \frac{c\gamma^2 - \xi(\varepsilon - \varepsilon_0)}{\varepsilon(\varepsilon - \varepsilon_0 - c\gamma^2/
   \varepsilon_0)},
    \label{eq26}
      \end{equation} 
and the energy integration limits: $\varepsilon_{imp-pr',-} = \varepsilon_{c,-}$ and $\varepsilon_{imp-pr',+} = \min[\varepsilon_{c,+},
\hbar\omega - \varepsilon_c]$. Then the function $\sigma_{imp-pr'}(\omega,T)$ follows from a numerical integration in Eq. \ref{eq23} and, as 
seen in  Fig. 3, it has a lower threshold frequency $\omega_c' = \varepsilon_c + \varepsilon_{c,-}$ than the $pr-pr'$-term. Above this threshold, 
it starts to grow linearly as $\sim (\omega/\omega_c'-1)c^{5/2}c_0^{-5/3}\sigma_0$ and, for the impurity concentrations within the "safety range", 
$c \ll c_1 \sim c_0^{2/3}$, becomes completely dominated by the $pr-pr'$-function, Eq. \ref{eq24} above its threshold $\omega_c$.

\begin{figure}[htbp]
\begin{center}
\includegraphics[width=8cm]{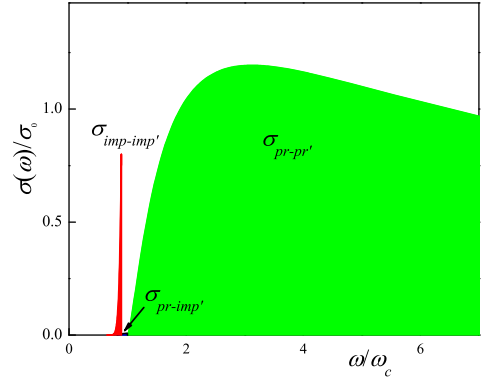}
\caption{General picture of the optical conductivity showing three types of contributions for the impurity perturbation parameters 
as chosen in Fig. \ref{fig1}.}
\label{fig3}
\end{center}
\end{figure}

Finally, the $imp-imp'$-term is obtained with a similar numerical routine on Eq. \ref{eq23}, using Eq. \ref{eq22} either for 
the poles $\xi_{1,2}(\varepsilon)$ by the $imp$-band and for the $\xi_{3,4}(\varepsilon')$ by the $imp'$-band and Eq. \ref{eq25} for respective 
generalized velocities while the energy integration limits in this case are $\varepsilon_{imp-imp',-} = \varepsilon_{c,-}$ and $\varepsilon_{imp-imp',+} = 
\min[\varepsilon_{c,+}, \hbar\omega - \varepsilon_{c.-}]$. The resulting function $\sigma_{imp-imp'}(\omega,T)$ occupies the narrow frequency band from 
$\omega_{imp-imp',-} = 2 \varepsilon_{c,-}/\hbar$ to $\omega_{imp-imp',+} = 2\varepsilon_{c,+}/\hbar$ (Fig. \ref{fig3}) and its asymptotics near these 
thresholds and in the zero-temperature limit are obtained analytically as:
\begin{equation}
   \sigma_{imp-imp'}(\omega,0) \approx \sigma_0\frac{16c^{7/2}\gamma^7}{3\sqrt 2 \xi_-^7}\left(\frac{\omega - \omega_{-}}{\omega_{-}}\right)^{3/2},
    \label{eq27}
      \end{equation} 
at $0 < \omega - \omega_{-} \ll \omega_{-}$ and a similar formula for $0 < \omega_+ - \omega \ll \omega_+$ only differs from it by the change: 
$\xi_-\to\xi_+$ and $\omega_-\to\omega_+$. 

Then the maximum contribution by the $imp-imp'$-term is estimated by extrapolation of the above asymptotics to the 
center of the impurity band: $|\omega - \omega_\pm| \sim |\omega_0 - \omega_\pm|$, resulting in: $\sigma_{imp-imp',max} 
\sim \sigma_0 c^5 c_0^{-10/3} (\xi_+/\xi_-)^{7/2}$. This estimate shows that the narrow $imp-imp'$-peak of optical conductivity 
around $\omega \approx 2\epsilon_0/\hbar$ can, unlike the "combined" $imp-pr'$-term, can become as intense or even more 
than the maximum of "principal" $pr-pr'$ intensity, Eq. \ref{eq24}, if the small factor $\sim (c/c_1)^5$ be overweighted by the 
next factor $(\xi_+/\xi_-)^{7/2}$. The latter is only possible if the impurity perturbation is {\it weak} enough: $v \ll 1$. Then the 
ratio $\xi_+/\xi_-$ turns $\approx (2/v)^2 \gg 1$ and can really overweight the concentration factor if the impurity concentration 
$c$ reaches $\sim c_1 (v/2)^{7/5} \ll c_1$, that is quite realistic within the "safety" range $c \ll 1$. The overall picture of optical 
conductivity for an example of weakly coupled, $v = 0.5$, impurities at  high enough concentration $c = 4c_0$ is 
shown in Fig. \ref{fig3}. The expressed effect of "giant" optical conductivity by the in-gap impurity excitations could be 
compared with the well known Rashba enhancement of optical luminescence by impurity levels at closeness to 
the edge of excitonic band \cite{rashba} or with the huge impurity spin resonances in magnetic crystals \cite{ilp}, 
but with a distinction that it appears here in a two-particle process instead of the above mentioned single-particle 
ones.

\section{Static kinetic coefficients}

Now we can pass to the relatively simpler calculation of the kinetic coefficients in the static limit of $\omega \to 0$. 
To begin with, consider the heat conductivity, Eq. \ref{eq16}, where the momentum integration at coincidence 
of the above mentioned poles $\xi_{1.3}$ and $\xi_{2.4}$ is readily done using the general convolution formula:
\begin{equation}
  \int L_{\Gamma_j}\left(\xi - \xi_j\right) L_{\Gamma_k'}\left(\xi- \xi_{k}'\right)d\xi = L_{\Gamma_j + \Gamma_k'}\left(\xi_j - \xi_k'\right),
    \label{eq28}
      \end{equation} 
for two Lorentzian fuctions $L_\Gamma(\xi) = \Gamma/(\xi^2 + \Gamma^2)$, and in the limit of $\xi_i = \xi_k'$ and $\Gamma_j = \Gamma_k'$ 
obtaining simply $(2\Gamma_j)^{-1}$, a "combined lifetime". This immediately leads to a Drude-like formula for heat 
conductivity as a sum of principal and impurity terms, $\kappa(T) = \kappa_{pr}(T) + \kappa_{imp}(T)$, each of them given by:
\begin{eqnarray}
  \kappa_{pr}(T) & = &  \frac{\hbar(1 + v^2)}{\pi cVv} \int_{\varepsilon_c}^{\infty} d\varepsilon \frac{\partial f(\varepsilon)}{\partial \varepsilon} \frac{\varepsilon\left(\varepsilon^2 
    - \varepsilon_0^2\right)}{\sqrt{\varepsilon^2 - \Delta^2}}\nonumber\\
& \approx & \frac{\hbar\rho_{\rm F}\Delta^2}{c} \sqrt{ \frac{\pi\beta\Delta}{2}}\exp (-\beta\Delta),
    \label{eq29}
      \end{eqnarray}
 and:
 \begin{eqnarray}
  \kappa_{imp}(T) & \approx &  \frac{\hbar}{\pi\left(\varepsilon_{c,+} - \varepsilon_{c,-}\right)} \left(\frac{c}{c_0}\right)^4\int_{\varepsilon_{c,-}}^{\varepsilon_{c,+}} 
   d\varepsilon \frac{\partial f(\varepsilon)}{\partial \varepsilon}\varepsilon^2\nonumber\\
& \approx & \frac{\hbar}{\pi} \left(\frac{c}{c_0}\right)^4 \beta\varepsilon_0^2\exp (-\beta\varepsilon_0).
    \label{eq29}
      \end{eqnarray}
Then the comparison of Eqs. \ref{eq28} and \ref{eq29} shows that the impurity contribution to the heat conductance 
$\kappa_{imp}$ for impurity concentrations $c$ above the critical value $c_0$ turns to dominate over the principal contribution 
$\kappa_{pr}$ at all the temperatures (of course, below the critical transition temperature). Such strong impurity effect is 
combined from enhanced thermal occupation of impurity states and from their growing lifetime as $\sim c^3$ against 
the decreasing as $\sim 1/c$ lifetime in the principal band. 

\begin{figure}[htbp]
\begin{center}
\includegraphics[width=8cm]{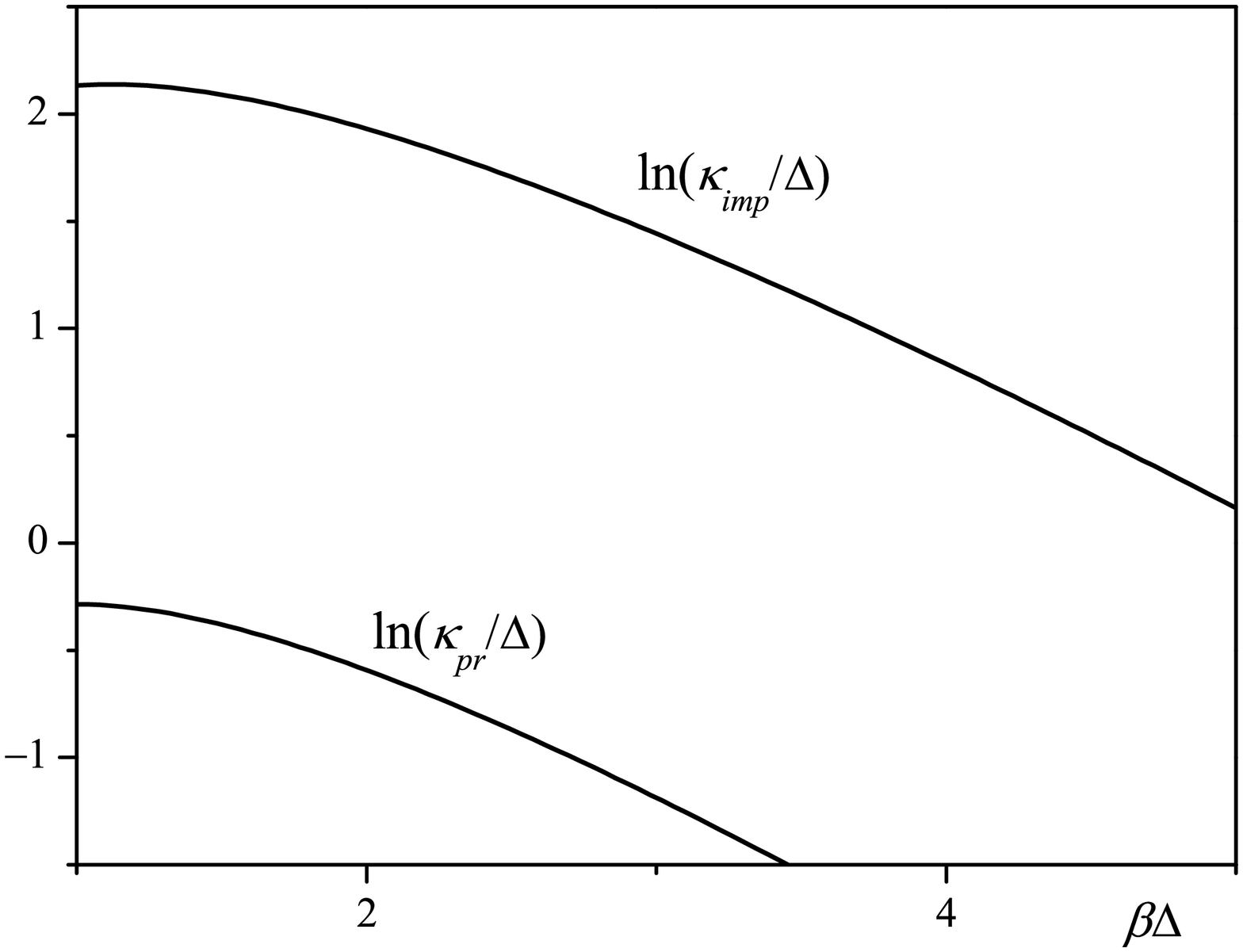}
\caption{Logarithmic plots for two contributions to the heat conductivity shows domination of the impurity term at all the 
temperatures where SC itself exists. The same impurity perturbation and concentration parameters are used as in Fig. \ref{fig1}}
\label{fig4}
\end{center}
\end{figure}

Similar strong impurity effects should also follow for the static electric conductivity $\sigma(0,T)$ (see \cite{note}) and for the 
thermoelectric Peltier and Seebeck coefficients, Eq. \ref{eq17}. All of them can be considered as fully due to the 
corresponding impurity contributions and the temperature dependencies of thermoelectric coefficients should be 
non-exponential: $\Pi(T) \approx \Pi(0) = $ const, and $S(T) \approx \Pi(0)/T$, alike the non-perturbed case but at much 
higher level. Finally, it is important to underline that the above predictions are only for impurity concentrations above 
the critical value, $c \gtrsim c_0$, while the system transport properties should stay almost non-affected by impurities 
below this concentration, $c < c_0$. Fig. \ref{fig4} demonstrates these differences between temperature dependencies of static 
conductivities and of thermoelectric coefficients for low and high concentrations of impurities at the choice of perturbation 
parameter as $v = 1$. Such drastic changes of transport behavior are of interest for experimental verification in properly 
prepared samples of SC ferropnictides with controlled concentration of specific impurities.

\section{Conclusive remarks}
In conclusion, the essential modification of quasiparticle spectra in a SC ferropnictide with impurities of simplest 
(local and non-magnetic) perturbation type is expected, consisting in formation of localized in-gap impurity states 
and their development into specific narrow bands of impurity quasiparticles at impurity concentration above a certain 
(quite low) critical value $c_0$ and leading to a number of effects in the system observable properties. Besides the 
previously discussed thermodynamical effects, expected to appear at all impurity concentrations, that is either due 
to localized or band-like impurity states, a special interest is seen in studying the impurity effects on electronic transport 
properties of such systems, only affected by the impurity band-like states. It was shown above that the latter effects can 
be very strongly pronounced, either for high-frequency transport and for static transport processes. In the first case, the 
impurity effect is expected to most strongly reveal in a narrow peak of optical conductance at its closeness to the edge 
of conductance band in non-perturbed crystal, resembling the known resonance enhancement of impurity absorption 
(or emission) processes near the edge of main quasiparticle band in normal systems, here it would be possible if the 
impurity perturbation be weak enough. The static transport coefficients at overcritical impurity concentrations are also 
expected to be strongly enhanced compared to those in a non-perturbed system, including the thermoelectric Peltier 
and Seebeck coefficients. The experimental verifications of these predictions would be of evident interest, since they 
can open perspectives for important practical applications, e.g., in narrow-band microwave devices or advanced 
low-temperature sensors, but this would impose rather hard requirements on the quality and composition of the necessary 
samples, they should be extremely pure aside the extremely low (by common standards) and well controlled contents 
of specially chosen and uniformly distributed impurity centers within the SC iron-arsenic planes of a ferropnictide 
compound. This situation can be compared to the requirements on doped semiconductor devices and hopefully should 
not be a real problem for modern lab technologies.

\section{Acknowledgements}
Y.G.P. and M.C.S. acknowledge the support of this work through the Portuguese FCT project PTDC/FIS/101126/2008.
V.M.L. is grateful to the EU Research Programs for a partial support through the grant FP7-SIMTECH No 246937.


\begin{thebibliography}{1}
\bibitem{note} Note that this static limit of Eq. \ref{eq13} only defines the conductivity by normal quasiparticles, seen e.g.
in normal resistivity by the magnetic flux flow in the mixed state, but otherwise short circuited by the infinite static conductivity
due to supercurrents.
\end{thebibliography}
\end{document}